
\magnification=1200
\global\newcount\meqno
\def\eqn#1#2{\xdef#1{(\secsym\the\meqno)}
\global\advance\meqno by1$$#2\eqno#1$$}
\global\newcount\refno
\def\ref#1{\xdef#1{[\the\refno]}
\global\advance\refno by1#1}
\global\refno = 1
\vsize=7.5in
\hsize=5.6in
\tolerance 10000

\baselineskip 18pt plus 1pt minus 1pt
\vskip 2in
\centerline{\bf FINE STRUCTURE DISCUSSION OF PARITY-NONCONSERVING}
\centerline{{\bf  NEUTRON SCATTERING AT EPITHERMAL ENERGIES}
\baselineskip 12pt plus 1pt minus 1pt
\footnote{*}{This work is
supported in part by funds
provided by the U. S. Department of Energy (D.O.E.)
under cooperative agreement
\#DE-FC02-94ER40818.}}
\vskip 24pt
\baselineskip 12pt plus 1pt minus 1pt
\centerline{M. S. Hussein
\footnote{$^{\dagger}$}{Permanent address: Instituto de F\'{\i}sica,
	      Universidade de S\~ao Paulo, C. P. 20516, S\~ao Paulo
	      S. P. Brazil. Supported partly by FAPESP (Brazil)},
	    A. K. Kerman and C-Y Lin
\footnote{$^{\ddagger}$}{Supported by Conselho de Desenvolvimento
			 Cient\'{\i}fico e Tecnol\'ogico(CNPq)}}
\vskip 12pt
\baselineskip 12pt plus 1pt minus 1pt
\centerline{\it Center for Theoretical Physics}
\centerline{\it Laboratory for Nuclear Science}
\centerline{\it and Department of Physics}
\centerline{\it Massachussets Institute of Technology}
\centerline{\it Cambridge, Massachusetts\ \ 02139\ \ \ U.S.A.}
\vskip 3.0cm
\vskip 24 pt
\baselineskip 15pt plus 2pt minus 2pt
\centerline{{\bf ABSTRACT}}
\medskip

	The large magnitude and the sign
correlation effect in the parity non-conserving resonant scattering
of epithermal neutrons from $^{232}$Th
is discussed in terms of a
non-collective $2p-1h$ local doorway model.
General conclusions are drawn as to the probability of
finding large parity violation effects in other regions of
the periodic table.

\vskip 24pt
\vfill
\noindent PACS numbers: 25.40.Ny, 11.30.Er, 24.10.Ht, 24.80.Dc

\noindent CTP\# 2296
\hfill September 08, 1994
\eject
\medskip
\baselineskip 18pt plus 2pt minus 2pt
\medskip
\nobreak
\xdef\secsym{}\global\meqno = 1
\medskip

	The discovery of sign correlation in parity
non-conserving (PNC) epithermal neutron-induced compound-nucleus
reactions involving the  heavy nucleus $^{232}{\rm Th}$ has
prompted intensive theoretical discussion concerning their origin.
The great interest in the TRIPLE data\ref\fret stems from the fact
that the statistical theory (ST) of these reactions,
although it predicts the possibility of large
PNC at individual resonances as the data also show, rules
out any sign correlations in the longitudinal asymmetry, contrary
to what the data show.

	The quantity usually analysed is the longitudinal asymmetry.
At a given $p1/2$ resonance in the compound nucleus
\eqn\nucle{
	P={\sigma_{R}^{(+)}-\sigma_{R}^{(-)}
	\over \sigma_{R}^{(+)}+\sigma_{R}^{(-)}}
	}
where $\sigma^{(\pm )}_{R}$ is the resonance neutron cross-section
for $\pm$ helicities at a given $p1/2$ resonance. We now use the usual formula
for the total cross-section
\eqn\trcss{
	\sigma_{T}^{(\pm)}
	={4\pi\over k} {\rm Im}f^{(\pm)}(0)
	=\sigma_{0}^{(\pm)}+\sigma_{R}^{(\pm)}
	   ,}
with
\eqn\witha{
	f^{(\pm)}(0)=f^{\pm}_0-{1\over 2k}\sum_{q}
	  {\gamma_{q}\bigl( \gamma_{q}\pm\gamma_{q}^{W}\bigr)
					     \over E-E_{q}+i\Gamma_{q}/2}
	}
where $\sigma^{\pm}_0$ is a background non-resonant piece
related to $f^{(\pm)}_{0}$, the latter being
roughly equal to $-a^{(\pm)}$ with $a^{(\pm)}$ being the scattering length,
$k$ is the asymptotic wave number, $E_{q}-i\Gamma_{q}/2$
is the complex energy of the ${q^{th}}$ resonance $(p1/2)$
in  the compound nucleus, $\Gamma_{q}$
is the total width (including $\gamma$-decay),
$\gamma_{q}$ is the strong component of the neutron decay amplitude
and  $\gamma_{q}^{W}$ is the ``weak'' component
of the neutron decay amplitude arising from coupling of the
$s1/2$ and $p1/2$ channels and  measures the PNC strength.
Thus one has
\eqn\thoha{
	P={\Delta\sigma_{R}\over 2\sigma_{R}}
	     ={\gamma^{W}\over\gamma}
	}
for each resonance and the average
$\langle P\rangle_{q}=\langle\gamma^{W}/\gamma\rangle_{q} $
was found to be about $0.08\pm 0.06$ (cf Ref.\fret\ ).
Theoretical attempts to explain the large value of
$\langle P\rangle_{q}$
have been made using different single particle
mechanisms\ref\bowm\ref\auer\ref\aubo\ref\caio. So far
no explanation has been put forth which does not require
on unrealistically large parity violating matrix element,
of about $100 eV$\ref\jjsz.\

	We discuss in what follows the fine structure aspect
of the data \ref\piza.\
A very natural mechanism that could account for this sign correlation
is to assume that the compound nuclear process
occurs through a {\it single} dominantly $p$-wave
{\it local} doorway which contains a small parity violation.
For simplicity we start with this extreme hypothesis. Below we will
consider the more general case including more doorways.
This doorway is a relatively simple
but statistical combination of two particle-one hole
$(2p-1h)$ states in $^{233}$Th.
In passing we note that
it is the dominance of local doorways which may give
rise to intermediate structure in the energy dependence
of nuclear cross-sections and their statistical nature which
gives rise to fluctuations in strength functions from nucleus to
nucleus over and above the general optical model trend\ref\feke\ref\fesh.
We stress that our local doorway states are statistical
in nature, in contrast to the collective $O^{-}$ doorway
(giant monopole) states considered by Auerbach\auer\aubo as responsible
for the sign correlation.
We assume that parity violation occurs through the coupling of our
$p$-wave doorway to
an $s$-wave doorway, located nearby. Then
\eqn\oftfe{
	\gamma_{q}=C_{qD}\gamma_{Dp}
	\qquad
	\gamma_{q}^{W}
	 =C_{qD}\gamma_{Ds}^{W}
	}
where $C_{qD}$ are taken to be random.
Taking $D$ to be in the vicinity of
the compound resonances in question, we obtain
with the aid of perturbation theory
\eqn\tlpri{
	 P
	= {\gamma_{Ds}^{W}\over\gamma_{Dp} }
	\simeq {\bar M^{W}\over\Delta E}{\gamma_{Ds}\over\gamma_{Dp}}
	,}
which represents the average value of $P$, being independent of $q.$
In Eq.\tlpri\ $\bar M^{W}$ is a characteristic
``weak'' matrix element
between $p$- and $s$-doorways and $\Delta E$
is the corresponding characteristic
energy distance between these doorway states
given by, e.g.,
\eqn\gibeg{
	\Delta E
	\simeq \Bigl| \bigl( E_{D_{2}} - i{\Gamma_{D_{2}}\over 2 }\bigr)
     - \bigl( E_{D_{1}} + i{\Gamma_{D_{1}}\over 2 }\bigr) \Bigr|
=\Bigl[ \bigl( E_{D_{2}}-E_{D_{1}}\bigr)^2 +
	{(\Gamma_{D_{1}}+\Gamma_{D_{2}})^2 \over 4}
 \Bigr]^{1/2}
	  }
In order to estimate the size of \tlpri\ we take the smooth energy dependence
out of the partial width amplitude $\gamma_{D}$ in the usual fashion,
i.e. we define
$\gamma_{Ds}/\gamma_{Dp}
=(1/kR)(\gamma_{Ds}^{(0)}/\gamma_{Dp}^{(0)})$, where $\gamma^{(0)}$
are the reduced widths. Thus
\eqn\witrw{
	\langle P \rangle
	={ \bar M^{W}\over \Delta E}
	 {\gamma^{(0)}_{Ds} \over \gamma^{(0)}_{Dp}}
	 {1\over kR}
	,}
which will have a definite sign for a given nucleus
as seen in the data.

	Properties of simple two particle-one hole states
can be deduced from the exiton model, usually employed in
pre-equilibrium studies\fesh. The density of $2p-1h$
doorway states coupled to total angular momentum $J$
at an excitation energy $E^{*}$ in the compound nucleus
is given by\fesh
\eqn\isgiv{
	\rho_{2p-1h} (E^{\ast},J)
	={g^{3} E^{\ast^{2}}\over 4}
	 {(2J+1)\exp [-(J+1/2) / 3\sigma^2] \over (27\pi)^{1/2}\sigma^3 }
	  }
where $g$ is the average spacing of single particle levels,
and $\sigma$ is the spin cut-off parameter. For the deformed
nucleus  $^{233}$Th, $g\sim 10$ $MeV^{-1}$, $\sigma\simeq 4.0$,
and taking $E^{*}=6MeV$, $J=1/2$, we find
$\rho_{2p-1h} \simeq 34 MeV^{-1}$.
Thus the average spacing, $D_{2p-1h}=(\rho_{2p-1h})^{-1} \simeq 30KeV.$
For a simple local doorway to dominate the
$2p-1h$ doorways must not be overlapping.
Thus we take $\Gamma\simeq D\sim 30KeV$.
Then $\Delta E$ of Eq.\gibeg becomes roughly
$\Delta E\sim \sqrt{2}\Gamma_{D} \sim 50 KeV$.

	Taking for $\bar M^{W}\sim 1.0 eV$\ref\fono
and $kR\sim 10^{-3}$
for $E_{n}\sim 1eV$, we find that the data
$(\langle P\rangle =0.08)$ require
$$
	\Bigl| {\gamma^{(0)}_{Ds}\over\gamma^{(0)}_{Dp}}\Bigr|
	\simeq 4.
$$
which seems to contradict the fact that p-waves are resonant
while s-waves are off resonance. This requires an enhancement which
may come about if the particular statistical doorway
involved in $\bar M^{W}$ couples strongly to $s$ and less so to $p$
or if the particular matrix element $\bar M_{W}$ is larger
than average.
This will be a statistical phenomenon associated with random
properties of the local doorways.

	The fluctuation part of $P$ can also be analysed
within the local doorway model.
If we consider ``nearby'' local doorways which for
simplicity we collectively call $D'$, then
\eqn\othen{
	\gamma_{q}=C_{qD}\gamma_{Dp}
		  +C_{qD'}\gamma_{D'p}
	\qquad
	\gamma_{q}^{W}
	 =C_{qD}\gamma_{Ds}^{W}
	 +C_{qD'}\gamma_{D's}^{W}
	.}
where
\eqn\where{
	C_{qD'} < C_{qD}
	}
on average. Then
\eqn\tlpro{
	 P\simeq {\gamma_{Ds}^{W}\over\gamma_{Dp} }
	+{C_{qD'}\over C_{qD}}
	\Bigl[ {\gamma_{D's}^{W}\over\gamma_{Ds}^{W}}
	-{\gamma_{D'p}\over\gamma_{Dp}}\Bigr]
	.}
The variance of $P$ is then given by
\eqn\tvigt{
	v\equiv
	\sqrt{\langle P^2\rangle-\langle P\rangle^2}
	\simeq
	\sqrt{\Bigl\langle \Bigl|
	{C_{qD'}\over C_{qD}}
	\Bigr|^2 \Bigr\rangle }
	\Bigl| {\gamma_{D's}^{W}\over \gamma_{Ds}^{W}}
	- {\gamma_{D'p}\over \gamma_{Dp}}\Bigr|
	.}
{}From Ref.\fret\ we find $v$ to be about unity. Because of \where\
\eqn\frfwf{
	\Bigl| {\gamma_{D's}^{W}\over \gamma_{Ds}^{W}}
	-{\gamma_{D'p}\over \gamma_{Dp}}\Bigr| > 1
	,}
which is entirely reasonable.
Thus the fine structure analysis furnishes us with constraining
relations involving the weak and strong decay amplitudes
of $D$ and $D'$.

	The presence of a dominant local doorway
that gives rise to a large value of
$\mid\gamma^{(0)}_{Ds}/\gamma^{(0)}_{Dp}\mid$
is certainly possible in some nuclei.
The probability that $\mid\gamma^{(0)}_{Ds}/\gamma^{(0)}_{Dp}\mid$
is, say, $f$ can be calculated
as follows. Since the local doorways are statistical in nature,
the $\gamma^{(0)}_{Ds}$ and $\gamma^{(0)}_{Dp}$ are Gaussian
distributed with about the same width. We thus have
\eqn\wthuh{\eqalign{
	P \Bigl( \bigl|{\gamma^{(0)}_{Ds}\over \gamma^{(0)}_{Dp}}\bigr|
	=f \Bigr)
	&={1\over \pi}\int_{-\infty}^{\infty} dx \int_{-\infty}^{\infty} dy
	 e^{-(x^2+y^2)} \delta \Bigl( \bigl|{x\over y}\bigr|-f \Bigr) \cr
	&={2\over\pi} {1\over 1+f^2}
		   .}}
Thus, within our model, the probability for the occurence of the phenomenon
of large parity violation {\it with} sign
correlation goes as $(1+f^2)^{-1}$, which is  small for
$f\sim 4$ which we found in n+$^{232}$Th. Therefore such a phenomenon
is not a global one that is exhibited by nuclei over the periodic table.
It happens in $^{232}$Th due to a statistical fluctuation
among the properties of local doorways.
In fact, in the n+$^{238}$U system,
also studied by the TRIPLE Group (Zhu et al,\fret), the average
value of $P$ was found to be zero and its magnitude smaller
so that $f \sim 1$.
This implies that a local dominant $2p-1h$ doorway state, so conspicuous
in $^{233}$Th, does not occur in $^{239}$U. With no single dominant doorway
present, e.g. if two closely spaced doorways are relevant in the energy
region of interest, the sign correlation disappears.

	Our analysis points to the conclusion that the
phenomenon of ``sign correlation'' is  purely a
conventional nuclear structure problem, and it is not connected
to ``exotics''. Further, the phenomenon occurs in Th
by statistical accident.
Before ending we mention that in Th the single particle
p-wave is resonant whereas the s-wave is not. Even so,
$f \simeq\mid\gamma^{(0)}_{Ds}/\gamma^{(0)}_{Dp}\mid$ was found to be about 4.
It seems to us that a more favorable case would be to have the p-wave
off-resonance and the s-wave on $(A\sim 170)$. This
would give rise to a larger $f$,
with higher probability. Of course, this will be experimentally
more difficult because the $p$-wave resonances in this region may be
too narrow.

\vfill
\eject
\centerline{\bf ACKNOWLEDGEMENT}
 	We thank J. D. Bowman and H. Feshbach for useful discussions.
\bigskip
\medskip
\nobreak
\medskip
\centerline{\bf REFERENCES}
\medskip
\nobreak
\medskip
\item{\fret} C. M. Frankle et al. {\it Phys. Rev. Lett.}
	     {\bf 67}, 564 (1991);
     	     C. M. Frankle et al. {\it Phys. Rev.}
	     {\bf C46}, 778 (1992).
	     X. Zhu et al. {\it Phys. Rev.}{\bf C46}, 768(1992);
	     J. D. Bowman et al. {\it Phys. Rev.}{\bf 48}, 1116(1993);
	    For a more recent review see J. D. Bowman et. al.,
	    Annu. Rev. Nucl. Part. Sci.{\bf 43}, 829 (1993).
\medskip
\item{\bowm} J. D. Bowman, G. T. Garvey, C. R. Gould, A. C. Hayes
	     and M. B. Johnson,
	     {\it Phys. Rev. Lett.}{\bf 68}, 780 (1992).
\item{\auer} N. Auerbach, {\it Phys. Rev.}{\bf C45}, R514 (1992).
\medskip
\item{\aubo} N. Auerbach and J. D. Bowman,
	     {\it Phys. Rev.}{\bf C46}, 2582 (1992).
\medskip
\item{\caio} C. H. Lewenkopf and H. A. Weidenm\"{u}ller,
	     {\it Phys. Rev.}{\bf C46}, 2601 (1992).
\medskip
\item{\jjsz} J. J. Szymanski, J. D. Bowman, M. Leuschner,
	    A. B. Brown and I. C. Girit,
	    {\it Phys. Rev.}{\bf C49}, 3297 (1994).
\medskip
\item{\piza} Fine structure analysis of processes involving discrete
	     symmetry breaking such as iso-spin were made by
	     A. F. R. de Toledo Piza and A. K. Kerman,
	     {\it Ann. Phys.}{\bf 48}, 173 (1968)
\medskip
\item{\feke} See, e.g.,  B. Block and H. Feshbach,
	     {\it Ann. Phys.} {\bf 23}, 47 (1963);
	     A. K. Kerman, L. S. Rodberg and J. E. Young,
	     {\it Phys. Rev. Lett.}{\bf 11}, 422 (1963);
	     H. Feshbach, A. K. Kerman and R. M. Lemmer,
	     {\it Ann. Phys.} {\bf 41}, 280 (1967).
\medskip
\item{\fesh} H. Feshbach, ``Theoretical Nuclear Physics: Nuclear Reactions''
	     (John Wiley and Sons, New York, 1992)
\medskip
\item{\fono} The exact size of these matrix elements needs further
	     detailed investigation using estimates of the two-body PNC force.
	     For a recent discussion see, e.g., A. C. Hayes
	     and I. S. Towner,
	     {\it Phys. Lett.}{\bf B302}, 157 (1993). See also\aubo
\medskip

\vfill
\end